\documentstyle[twoside,fleqn,espcrc2,epsfig]{article}

\title{ \vspace{-.35in} Thermodynamics of Lattice QCD with massless quarks 
         and chiral 4-fermion interactions.\thanks{Talk presented by 
         D.~K.~Sinclair at LATTICE'98, University of Colorado, Boulder, 
         Colorado.}}

\author{J.~B.~Kogut\address{Department of Physics, University of Illinois,
        1110 West Green Street, Urbana, IL 61801, USA},
        J.-F.~Laga\"{e}$^{\rm b}$ and D.~K.~Sinclair\address{HEP Division, 
        Argonne National Laboratory, 9700 South Cass Avenue, Argonne, 
        IL 60439, USA}}

\begin{document}

\begin{abstract}
We have simulated lattice QCD with an irrelevant 4-fermion interaction and
2 zero mass quarks. The chiral phase transition is observed to be second order
and we discuss extraction of critical exponents.
\end{abstract}

\maketitle

\section{The action}

We have extended the standard staggered quark QCD action by the addition of a
4-fermion term with a $U(1) \times U(1)$ chiral flavour symmetry, generated
by $(\xi_5,\gamma_5\xi_5)$ \cite{kls}. (For earlier, related work see 
\cite{early}.) Introducing the auxiliary fields $(\sigma,\pi)$ to render the
action quadratic we have
\begin{eqnarray}
L\!\! & = &\!\!\! -\beta\sum_{\Box}[1-\frac{1}{3}{\rm Re}({\rm Tr}_{\Box} UUUU)]
        +{N_f \over 8}\sum_s \dot{\psi}^{\dag} A^{\dag} A\dot{\psi} \nonumber\\
 \!\! &   &\!\!\! -\sum_{\tilde{s}}\frac{1}{8}N_f\gamma(\sigma^2+\pi^2)   
        +\frac{1}{2}\sum_{\tilde{s}}(\dot{\sigma}^2+\dot{\pi}^2)    \nonumber\\ 
 \!\! &   &\!\!\! +\frac{1}{2}\sum_l(\dot{\theta}_7^2+\dot{\theta}_8^2
        +\dot{\theta}_1^{\ast}\dot{\theta}_1
        +\dot{\theta}_2^{\ast}\dot{\theta}_2
        +\dot{\theta}_3^{\ast}\dot{\theta}_3)
\end{eqnarray}
for our molecular dynamics lagrangian, where
\begin{equation}
A = \not\!\! D + m + \frac{1}{16} \sum_i (\sigma_i+i\epsilon\pi_i)
\end{equation}

The advantage of this action is that the Dirac operator remains non-singular
when $m=0$ so that we can work in the chiral limit. Note that we have an
exact chiral symmetry generated by $\gamma_5\xi_5$ which means we will have
a massless pion at $m=0$ when this symmetry is spontaneously broken.

The 4-fermion term is an irrelevant operator, so this theory should have the
same continuum limit --- continuum QCD --- as the standard action.

\section{Application to $N_f=2$ thermodynamics}

$N_f=2$ QCD is believed to have a second order finite temperature chiral phase
transition for $m=0$ which weakens to a rapid crossover, with no actual phase
transition, for $m \neq 0$. The critical exponents that describe the 
neighbourhood of this critical point are expected to be those of an $O(4)$
spin model in 3 dimensions. (See \cite{laermann} for a recent review of lattice
QCD thermodynamics and critical exponents.)

For the staggered lattice formulation the critical exponents are expected to be
$O(2)$ or $O(4)$. (For recent work on staggered critical exponents see 
\cite{milc}.) Here the advantage of being able to work at zero quark mass
is clear. We can hope to see clear evidence for the second order nature of
the phase transition. In addition one can measure certain critical exponents
directly. In particular one can measure the exponent $\beta_m$ from
\begin{equation}
\langle\bar{\psi}\psi\rangle = const (\beta_c - \beta)^{\beta_m}
                             = \gamma \langle\sigma\rangle
\end{equation}
where $\beta_m=0.5$ for mean field theory, $\beta_m \approx 0.38$ for $O(4)$,
and $\beta_m \approx 0.35$ for $O(2)$. In the high temperature phase, the
common $\pi$ and $\sigma$ screening masses increase from zero as
\begin{equation}
m^2_{\pi/\sigma} = const (\beta - \beta_c)^{\gamma_m}
\end{equation}
where $\gamma_m = 1$ for mean field theory, $\gamma_m \approx 1.44$ for $O(4)$
and $\gamma_m \approx 1.32$ for $O(2)$. The same critical exponent governs the
vanishing of the $\sigma$ mass below the transition. If we now fix 
$\beta=\beta_c$ and decrease the mass $m$ from a finite value to zero,
$\langle\bar{\psi}\psi\rangle$ vanishes as
\begin{equation}
\langle\bar{\psi}\psi\rangle = const \; m^{1/\delta}
\end{equation}
where $\delta = 3$ for mean field theory, $\delta \approx 4.8$ for $O(4)$, and
$\delta \approx 4.8$ for $O(2)$.

Of the fluctuation quantities (susceptibilities), the most accessible is the
``magnetic'' susceptibility
\begin{equation}
\chi_\sigma = {1 \over V} [ \langle (\sum_x \sigma(x))^2 \rangle
                          - \langle (\sum_x \sigma(x)) \rangle^2 ]
\end{equation}
which scales as 
\begin{equation}
\chi_\sigma = const (\beta - \beta_c)^{-\gamma_m}.
\end{equation}

\section{Lattice simulations and preliminary results}

Our simulations are being carried out using hybrid molecular dynamics with
noisy fermions to accommodate $N_f=2$. Because of the exact flavour 
$U(1)_{axial}$ symmetry at $m=0$, the direction of symmetry breaking in the
($\langle\bar{\psi}\psi\rangle$,$i\langle\bar{\psi}\gamma_5\xi_5\psi\rangle$)
space (or ($\sigma$,$\pi$) space is not determined. On a finite lattice, this
direction rotates during the run, forcing us to use
$\sqrt{\langle\bar{\psi}\psi\rangle^2
-\langle\bar{\psi}\gamma_5\xi_5\psi\rangle^2}$ or 
$\sqrt{\langle\sigma\rangle^2+\langle\pi\rangle^2}$ as our order parameter on
each configuration. (Here $\langle\rangle$ should be taken to mean a lattice
average, not an ensemble average.) This estimate differs from the true value by 
${\cal O}(1/\sqrt{V})$, which can only be removed by working at more than 1
spatial volume for each $N_t$.

We are currently running on $8^3 \times 4$, $12^2 \times 24 \times 4$,
$12^3 \times 6$, and $18^3 \times 6$ lattices at $\gamma = 20$, and on a
$12^3 \times 6$ at $\gamma = 10$.

\begin{figure}[htb] 
\epsfxsize=3in
\centerline{\epsffile{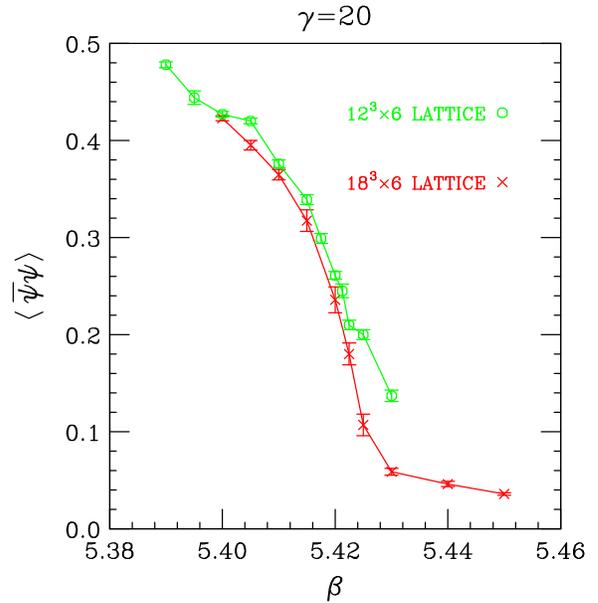}}
\vspace{-0.2in}
\caption{$\langle\bar{\psi}\psi\rangle$ as a function of $\beta$ for $N_t=6$
and $\gamma=20$.\label{fig:pbp}}
\vspace{-0.1in}
\end{figure}

\begin{figure}[htb] 
\epsfxsize=3in
\vspace{-0.1in}
\centerline{\epsffile{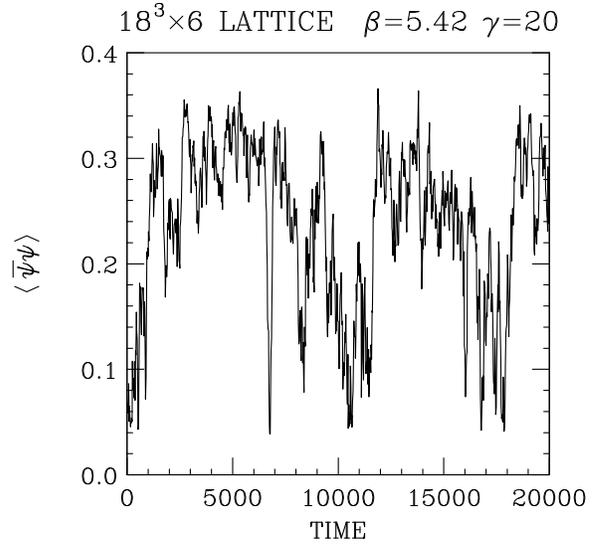}}
\vspace{-0.1in}
\caption{Time evolution of $\langle\bar{\psi}\psi\rangle$ at $\beta=5.42$ on an
$18^3 \times 6$ lattice at $\gamma=20$.\label{fig:pbp_time}}
\vspace{-0.1in}
\end{figure}

Whereas our earlier simulations at $N_t=4$ and $\gamma=10$ showed a first order
transition, all $N_t=6$ combinations above show evidence for the 
expected second order transition. $\langle\bar{\psi}\psi\rangle$ and the
Wilson line show a sharp, but not discontinuous, transition with no sign of
metastability, unlike the previous case. Figure~\ref{fig:pbp} shows the
$\beta$ dependence of the chiral condensate for $N_t=6$ and $\gamma=20$.
In addition, close to the transition, observables show clear signs of critical
slowing down with large fluctuations over many thousands of time units. (We
have simulated for as many as 39,000 time units at a single $\beta$ value.)
Such a time evolution for the chiral condensate close to the transition is
shown in figure~\ref{fig:pbp_time}.
The $N_t=4$, $\gamma=20$ runs show signs of critical fluctuations, but the
transition occurs over a very narrow range and it will require more statistics
to tell if it has become second order or remains first order.

It is clear that we will need to extend our runs close to the phase transition.
This is a consequence of the critical slowing down close to the second order
transition. This currently limits our ability to extract accurate critical
exponents due in part to the fact that our errors are poorly determined and
partly because it prevents accurate extrapolation to infinite volume. In
addition, the scaling window could be very narrow requiring us to consider
more $\beta$ values close to the transition. Despite this the
$\langle\bar{\psi}\psi\rangle$ data looks promising, and suggests a critical
exponent $\beta_m \sim 0.3$, however, this is on the basis of statistically
poor fits. The $N_t=6$, $\gamma=20$ data shows relatively modest finite volume
effects.

For $N_t=4$, $\gamma=20$, we find $\beta_c \approx 5.288$. For $N_t=6$, 
$\gamma=10$, $\beta_c \approx 5.466$, while for $N_t=6$, $\gamma=20$,
$\beta_c \approx 5.424$.

\begin{figure}[htb] 
\epsfxsize=3in
\centerline{\epsffile{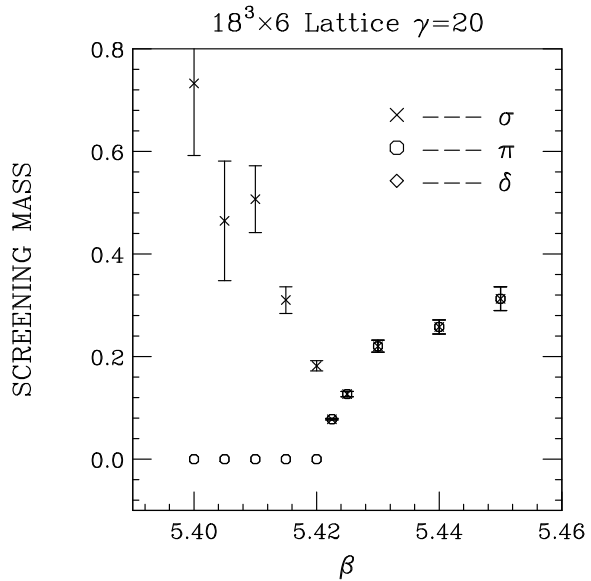}}
\vspace{-0.2in}
\caption{$\sigma$ and $\pi$ screening masses on an
$18^3 \times 6$ lattice at $\gamma=20$.\label{fig:masses}}
\vspace{-0.1in}
\end{figure}

Figure~\ref{fig:masses} shows the $\sigma$ and $\pi$ screening masses for the
$18^3 \times 6$, $\gamma=20$ simulations. The $\sigma$ screening mass shows
the correct behaviour, decreasing to zero as $\beta$ is increased through the
transition, and increasing from zero, degenerate with the $\pi$ above the
transition. Below the transition, the pion mass is consistent with zero. It
appears doubtful that these masses can be determined accurately enough to
obtain a good estimate for $\gamma_m$. The $\delta$ mass can be calculated,
but even assuming it remains distinct from the $\sigma/\pi$ mass above the
transition, it will be difficult to tell how much of this is due to the
explicit $U(1)_{axial}$ symmetry breaking provided by the 4-fermion
interaction.

\section{Summary}

$N_f=2$ lattice QCD with massless quarks and a weak 4-fermion interaction
appears to have the expected second order transition, at least for $N_t \geq 6$.
More work is needed to clarify the $N_t=4$ case.

With more statistics the $N_t=6$ simulations should produce an accurate
determination of the critical exponent $\beta_m$.
Moving to finite mass at $\beta=\beta_c$ should allow an accurate determination
of $\delta$.

Hadronic screening masses need further analysis. Other order parameters remain
to be analysed.

Unfortunately, there is no obvious way to include 4-fermion interactions with 
full $SU(2) \times SU(2)$ chiral flavour symmetry.

\section*{Acknowledgements}
Supported by DOE contract W-31-109-ENG-38 and NSF grant NSF-PHY-96-05199.
Computing was provided by NERSC.

\end{document}